\documentclass[useAMS,usenatbib,fleqn,letter]{mn2e}
\usepackage{graphics,graphicx,times}
\usepackage{amsmath}
\usepackage{amssymb}
\usepackage{aas_macros}
\usepackage[usenames,dvipsnames]{xcolor}

\newcommand{\sigmasfr}{\dot \Sigma_{\rm \star}}
\newcommand{\csigmasfr}{\dot\Sigma_{\rm \star,0}}
\newcommand{\sigmasfrcrit}{\dot\Sigma_{\rm \star,crit}}
\newcommand{\eagle}{{\sc eagle}}

\usepackage{color}
\usepackage[para]{footmisc}

\title[Brighter galaxies reionised the Universe]{The brighter galaxies reionised the Universe} 
\author[Sharma et al.]
{Mahavir Sharma$^1$\thanks{mahavir.sharma@durham.ac.uk}, Tom Theuns$^1$, Carlos Frenk$^1$, Richard Bower$^1$, Robert Crain$^2$, \newauthor 
Matthieu Schaller$^1$ \& Joop Schaye$^3$\\
$^{1}$ Institute for Computational Cosmology, Department of Physics, University of Durham, South Road, Durham, DH1 3LE, UK\\
$^{2}$ Astrophysics Research Institute, Liverpool John Moores University, 146 Brownlow Hill, Liverpool L3 5RF, UK\\
$^{3}$ Leiden Observatory, Leiden University, P.O. Box 9513, 2300 RA Leiden, the Netherlands
}

\begin{document}
\date{\today}
\pagerange{\pageref{firstpage}--\pageref{lastpage}} \pubyear{2014}
\maketitle

\begin{abstract}
Hydrogen in the Universe was (re)ionised between redshifts $z\approx 10$ and $z\approx 6$. The nature of the sources of the ionising radiation is hotly debated, with faint galaxies below current detection limits regarded as prime candidates. Here we consider a scenario
 	in which ionising photons escape through channels punctured in the interstellar medium by outflows powered by starbursts. 
 	We take account of the observation that strong outflows occur only when
the star formation density is sufficiently high, and estimate the
galaxy-averaged escape fraction as a function of redshift and luminosity
from the resolved star formation surface densities in the EAGLE
cosmological hydrodynamical simulation.
	We find that the fraction of ionising photons that escape from galaxies increases rapidly with redshift, reaching values of 5-20~per cent at $z>6$, with the brighter galaxies having higher escape fractions. Combining the dependence of escape fraction on luminosity and redshift with the observed luminosity function, we demonstrate that galaxies emit enough ionising photons to match the existing constraints on reionisation while also matching the observed UV-background post-reionisation. Our findings suggest that galaxies above the current {\em Hubble Space Telescope} detection limit emit half of the ionising radiation required to reionise the Universe.
  \end{abstract}


\section{Introduction}
Consensus is emerging that neutral hydrogen in the Universe was (re)ionised between redshifts $z\approx 10$ and $z\approx 6$ \citep[e.g.][]{Robertson15,McGreer15,Mitra15} . The nature of the sources of the ionising radiation has not yet been firmly established, but attention has focussed on an early generation of galaxies. Regions around such galaxies are ionised first, and these ionised bubbles grow in number and size until they percolate as more and brighter galaxies form \citep{Gnedin00,Shin08}. Therefore, as reionisation proceeds, a larger fraction of the Universe becomes ionised (see e.g. \citet{Loeb01} for a review). 

A crucial factor in modelling reionisation is the fraction of ionising photons that escape their natal galaxy, $f_{\rm esc}$. Neutral hydrogen and dust in the interstellar medium (ISM) and circumgalactic medium (CGM) determine a galaxy's $f_{\rm esc}$. This quantity is difficult to measure and observational claims of the detection of ionising photons escaping from galaxies are controversial.  The value of $f_{\rm esc}$ is often assumed to be constant \citep[e.g.][]{Bouwens11,Robertson13} or increasing towards lower luminosities \cite[e.g.][]{Ferrara13}. In such models, faint galaxies far below current detection limits dominate the ionising emissivity.

Observed values of $f_{\rm esc}$ at $z\approx 0$ are generally a few per cent or less. For example, \cite{Bland-Hawthorn01} infer a value of $f_{\rm esc}\approx 1-2$ per cent for the Milky Way galaxy. Such low values are not surprising since star forming regions are usually enshrouded in neutral gas with column density $N_{\rm H{\sc I}}\gtrsim 10^{20}$~cm$^{-2}$, three orders of magnitude higher than the value that yields an optical depth of $\tau=1$ for ionising photons with energy 1 Rydberg; it is this neutral gas that fuels star formation in the first place. For galaxies to be able to reionise the Universe by $z\approx 6$ and provide the bulk of the ionising photons post-reionisation, $f_{\rm esc}$ needs to increase rapidly with redshift, $\propto (1+z)^{3.4}$ according to \citet{Haardt12} or even faster \citep{Khaire15}. Such a rapid evolution of the escape fraction then suggests that the ISM/CGM of high-$z$ galaxies is fundamentally different from that at low $z$; the expected decrease in dust content is not enough to explain the trend. Why this should be so is a mystery whose resolution is key to understanding cosmological reionisation.

Whether the escape fraction indeed evolves (rapidly) can in principle be tested directly by measuring $f_{\rm esc}$ as function of redshift. There are currently no direct detections of ionising photons escaping from individual galaxies at $z\approx 1$ \citep{Bridge10,Siana10,Rutkowski15}, with 3~$\sigma$ upper limits of $f_{\rm esc}\approx 2$~per cent. In contrast, \cite{Nestor13} measure $f_{\rm esc}=5-7$~per cent for $z\approx 3$ Lyman-break galaxies  \citep[see also][]{Vanzella12} and even higher values for Lyman-$\alpha$ emitters. These values are significantly higher than more recent determination of a value $< 2$~per cent by \cite{Grazian15} for LBGs at $3.3<z<4$. The observational evidence for evolution in $f_{\rm esc}$ is thus currently inconclusive.

Numerical simulations of galaxy formation that include radiative transfer, aimed at calculating $f_{\rm esc}$ at $z\gtrapprox 6$, also yield contradictory results. \cite{Kimm14} quote values of $f_{\rm esc}\approx 10$~per cent, with even higher values of $\approx 20$~per cent during starbursts. \cite{Paardekooper15} find much lower values, and claim that $f_{\rm esc}$ decreases rapidly with halo mass and cosmic time (see also \citealt{Yajima11,Razoumov09,Wise14}). In contrast, \cite{Ma15} find intermediate values, $f_{\rm esc}\approx 5$~per cent, with no strong dependence on either galaxy mass or cosmic time. Most ionising photons that do not make it out of their natal galaxy are absorbed locally - within several tens of parsecs of the star that emitted them. Consequently, the level of discrepancy between current simulations is not surprising, since the detailed properties of the ISM need to be modelled very accurately, with little guidance from observations.

A generic feature of these radiative transfer simulations is that starbursts clear channels in the ISM through which ionising photons escape \citep[e.g.][]{Fujita03}, a phenomenon that is particularly efficient at $z\gtrapprox 6$ when galaxies are particularly bursty \citep[e.g.][]{Wise08} and drive winds. { \cite{Heckman01} uses X-ray and optical emission line data to
	conclude that winds occur in galaxies with star formation surface density above a critical value of $\sigmasfrcrit\approx 0.1$~M$_\odot$~yr$^{-1}$~kpc$^{-2}$. Theoretical models of outflows driven by starbursts through supernovae \citep{Clarke02}, radiation pressure \citep{Murray11} or turbulent stirring \citep{Scannapiecco12}, support the existence of such a threshold, at similar values of $\sigmasfrcrit$ as inferred from observations.

If the lifetime of massive stars is shorter than the time required to create channels in the ISM, then the fraction of all ionising photons that escapes from a galaxy may be small, even though $f_{\rm esc}$ is high once the channels have been created \citep{Kimm14}. Compact starbursts may suffer less from such a timescale mismatch. Indeed, a wind moving with speed $v\geq 100$~km~s$^{-1}$ can carve a $500$~pc wide channel in a time $\leq 5$~Myr, comparable to the typical lifetime of a massive star. The few observed cases with large emissivities indeed result from very compact starbursts. \cite{Borthakur14} and \cite{Izotov16} find $f_{\rm esc}\approx 20$ and 8~per cent, respectively, in $z\approx 0.2-0.3$ compact starbursts; \cite{deBarros15} find a relative escape fraction of 60~per cent in a $z\approx 3$ compact starburst, and there is tantalising evidence that photons escape in channels \citep[e.g.][]{Chen07,Zastrow13}.

Here we propose a model in which $f_{\rm esc}$ for a star forming region depends on the local star formation surface density averaged on a scale of $\approx$1~kpc$^2$, $\sigmasfr$, motivated by the arguments presented above. We assume that when $\sigmasfr \geq  \sigmasfrcrit$ the escape fraction, $f_{\rm esc}$, is constant and equal to $f_{\rm esc,max}$ due to self-regulation of star-formation. We take $f_{\rm esc,max}\approx0.2$, roughly the observed upper limit, but explore how our results change if $f_{\rm esc,max}$ is varied between $0.1$ and $0.4$. When $\sigmasfr<\sigmasfrcrit$ we take $f_{\rm esc}=0$. We use this {\em Ansatz} to derive $f_{\rm esc}$ for all galaxies in the \eagle\ simulation \citep{Schaye15, Crain15}. We do not perform radiative transfer on \eagle\ galaxies directly since these simulations do not have enough resolution to model the physics that generates outflow driven channels.}

In Section~2, we calculate how the escape fraction depends on luminosity and redshift, illustrate what this implies for reionisation, the amplitude of the ionising background below $z=6$, and the nature of the galaxies that reionised the Universe. We summarise in Section~3.

\section{Escape of ionising photons from galaxies}
\subsection{The evolution of the star formation surface density}
\label{sec_SFR}

\begin{figure}
 \centering
 \includegraphics[width=1.\linewidth]{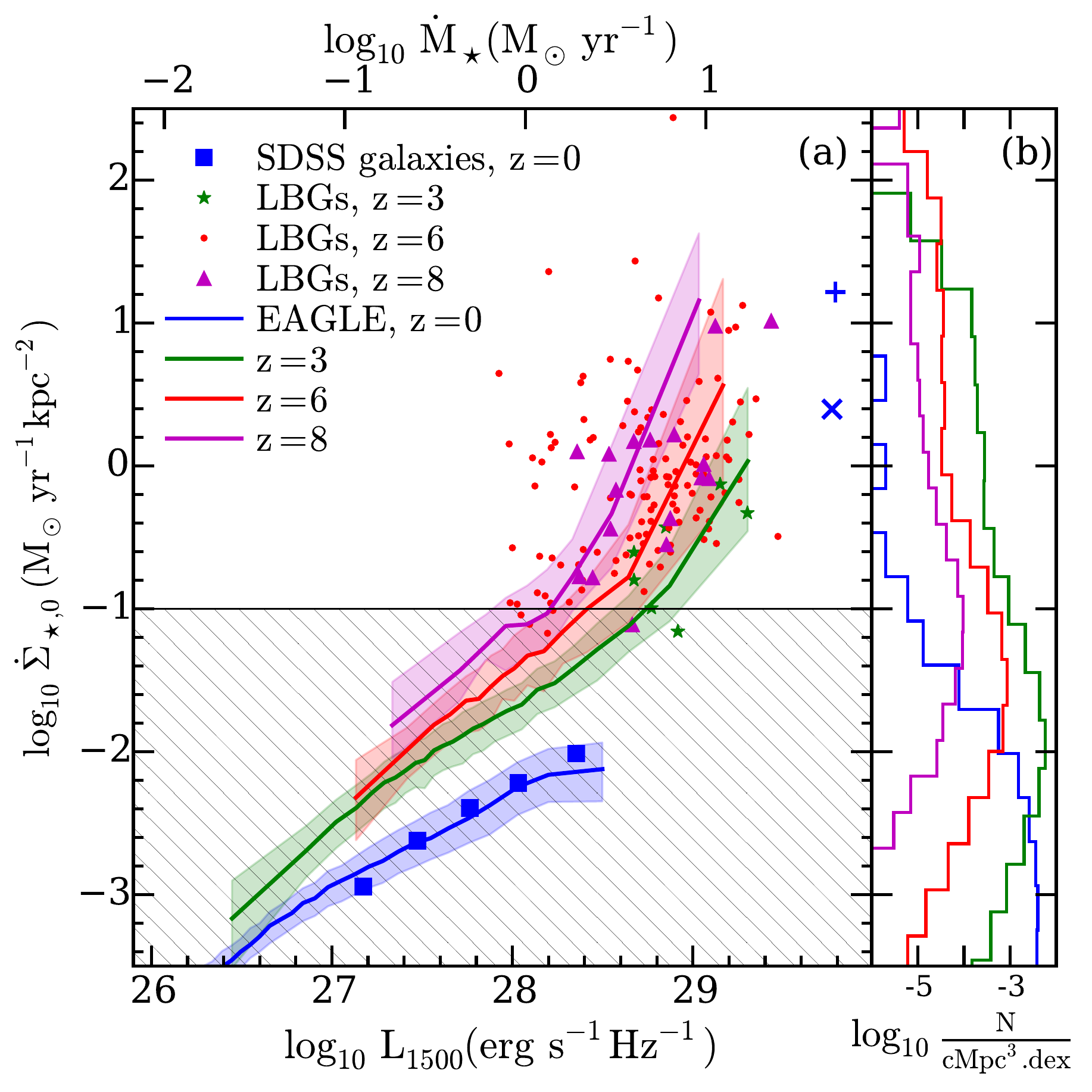}
\caption{The dependence of the mean surface density of star formation, $\csigmasfr$, on star formation rate and redshift.  {\bf  Panel a:} $\csigmasfr$ of \eagle\ galaxies with stellar mass $>10^8$ M$_\odot$, as a function of their 1500\AA\ UV luminosity ($L_{1500}$, bottom axis, corresponding star formation rate, $\dot M_\star$, top axis), with coloured lines depicting the median relation, and coloured regions enclosing the 25th to 75th percentile range. Different colours refer to different redshifts ({\color{blue} {\em blue}}, {\color{blue} $z=0$}, {\color{green} {\em green}}, {\color{green} $z=3$}, {\color{red} {\em red}}, {\color{red} $z=6$}, {\color{magenta} {\em magenta}}, {\color{magenta} $z=8$}). $\csigmasfr$ increases with $\dot M_\star$ at given $z$, and with $z$ at given $M_\star$. Horizontal black line shows the critical star formation rate surface density ($\sigmasfrcrit$) for outflows \citep{Heckman01}.  {\bf  Panel b:} Number density of \eagle\ galaxies in bins of $\csigmasfr$, using the same colour scheme. At $z=0$, galaxies with $\csigmasfr\gtrsim\sigmasfrcrit$ above which significant winds develop  are present in the simulation, but they are very rare. However at $z\gtrsim 6$, most \eagle\ galaxies with $\dot M_\star>1$~M$_\odot$~yr$^{-1}$ are above this limit, and are expected to drive strong winds. Observed values of $\csigmasfr$ are plotted in panel a, with colours depending on $z$ matching those of the simulation; the region of low $\csigmasfr$ where winds are not expected is hashed.  The data shown are: $z=0$, {\em Sloan Digital Sky Survey} measurements \protect\citep{Brisbin12}
    ({\color{blue} squares}), the galaxies with high values of $f_{\rm esc}$ from \protect\cite{Borthakur14} ({\color{blue} cross}) and \protect\cite{Izotov16} ({\color{blue} plus}), Lyman-break galaxies from \citet{Giavalisco96} at $z=3$ ({\color{green} stars}), from \citet{Curtis14} at $z=6$  ({\color{red} small dots}) and at $z=8$ ({\color{magenta} triangles}). \eagle\ galaxies reproduce the dependence of $\csigmasfr$ on $\dot M_\star$ and $z$.}
 \label{fig_LAsfr}
\end{figure}

\begin{figure}
 \centering
 \includegraphics[width=\columnwidth]{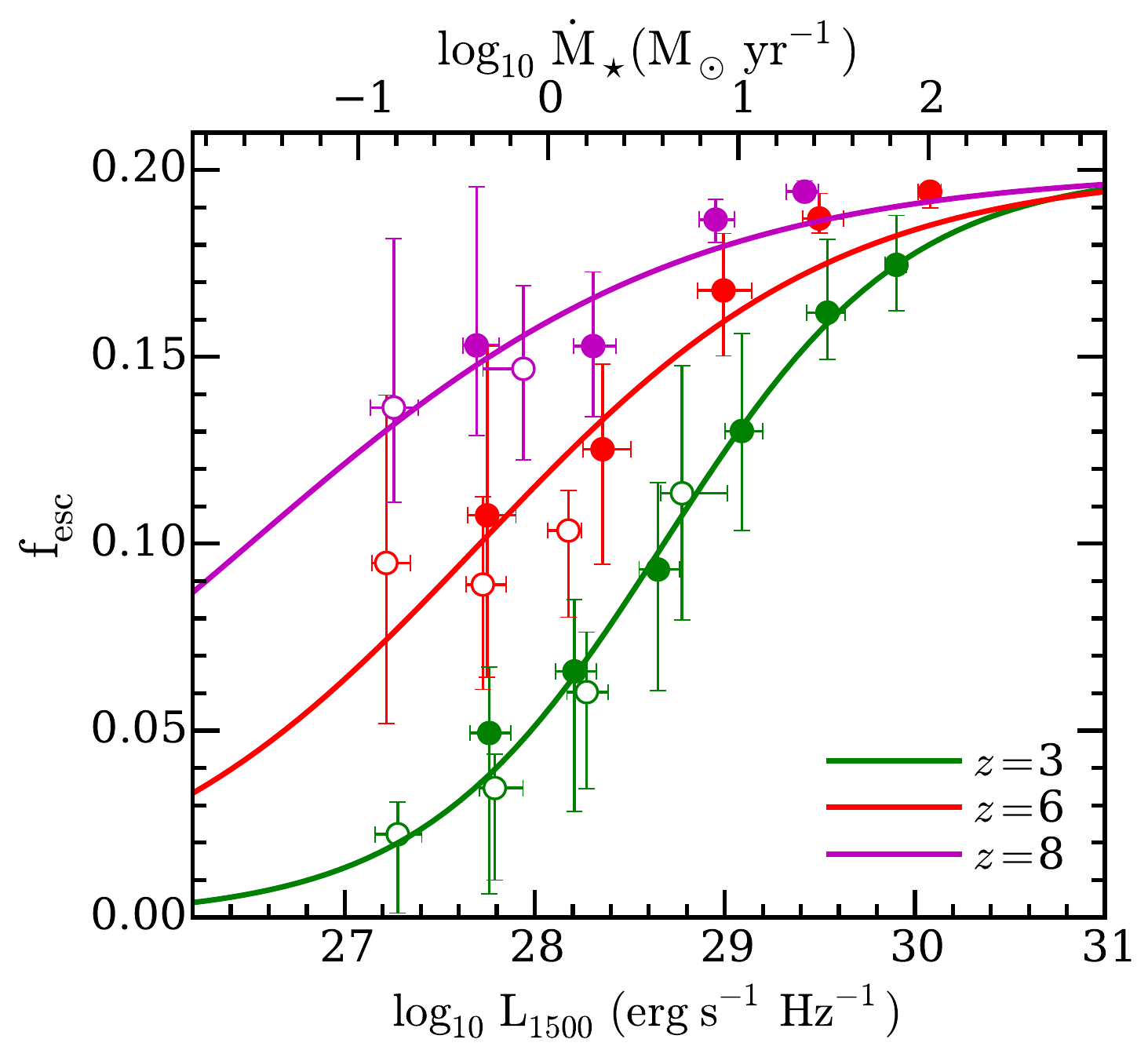}
 \caption{Mean inferred escape fraction for ionising photons ($f_{\rm esc}$) from galaxies in \eagle\ as a function of the 1500~\AA\ luminosity ($L_{1500}$) at three redshifts (green: $z=3$, red: $z=6$, magenta: $z=8$), for $f_{\rm esc, max}=0.2$. Circles represent the luminosity-weighted average of $f_{\rm esc}$, with error bars encompassing the 25th to 75th percentile range. Filled circles refer to the reference simulation (labelled Ref-L0100N1504), open circles to the simulation that has eight times better mass resolution (labelled Recal-L0025N0752). Simulation results are fitted with a sigmoid function shown as solid curves. The escape fraction increases with luminosity and redshift.} 
\label{fig_fesc}
\end{figure}
The \eagle\ hydrodynamical cosmological simulations \citep{Schaye15, Crain15} use subgrid modules for star and black hole formation, and feedback from stars and AGN. The star formation rate is calculated as function of pressure that ensures that galaxies follow the observed relation at $z=0$ between gas surface density and $\sigmasfr$ from \cite{Kennicutt98}, as described in \cite{Schaye08}, and the model assumes that this relation does not evolve. The star formation rate is assumed to be zero below the metallicity-dependent threshold of \cite{Schaye04}. Gas particles are stochastically converted into star particles with a probability per unit time that depends on their star formation rate. The \eagle\ suite includes models that vary in numerical resolution to test for convergence.  Here we use two models, Ref-L0100N1504 (linear size $L=100$~Mpc, gas particle mass $m_g=1.81\times10^6$ M$_\odot$) and the higher resolution model Recal-L0025N0752 ($L=25$~Mpc, $m_g=2.26\times10^5$ M$_\odot$).

 We compute $f_{\rm esc}$ of a galaxy by calculating the escape fraction of all its individual star-forming regions, and
	weighting them by their star formation rate. For an individual region we take $f_{\rm esc}=f_{\rm esc,max}$ when $\sigmasfr\geq\sigmasfrcrit$ and zero otherwise. Galaxies in \eagle\ are both more active (higher $\dot M_\star/M_\star$) and smaller at earlier times. As a consequence, an increasing fraction of star-forming regions have higher ISM pressure (see Fig.~7 in \citealt{Crain15}), which have higher $\sigmasfr$ according to the star formation prescription of \cite{Schaye08}, and hence higher $f_{\rm esc}$ as well.

The local quantity $\sigmasfr$ cannot be compared directly to observations. We therefore compute $\csigmasfr \equiv \dot M_\star/ (2\pi\,R_\star^2)$, where $\dot M_\star$ is the star formation rate of the galaxy and $R_\star$ its half-mass radius; $\csigmasfr$ is the central surface density of star formation if the disk is exponential. In \eagle, $\csigmasfr$ increases with $\dot M_\star$ at given redshift, and with redshift at given $\dot M_\star$ (coloured bands in Fig.~\ref{fig_LAsfr}). For example, for a galaxy with $\dot M_\star=1$~M$_\odot$~yr$^{-1}$, $\csigmasfr=10^{-2}$~M$_\odot$~yr$^{-1}$~kpc$^{-2}$ at $z=0$, but is more than a factor of 10 higher at $z=8$. The histogram in Fig.~\ref{fig_LAsfr} shows this more quantitatively. 

Observed galaxies display a very similar dependence of $\csigmasfr$ on $\dot M_\star$ and $z$ as \eagle\ galaxies (Fig.~\ref{fig_LAsfr}, data points are $\csigmasfr$ as function of $\dot M_\star$ for observed galaxies, with colours depending on redshift in the same manner as the coloured bands representing \eagle). That \eagle\ reproduces these trends so well is consistent with the fact that the simulation reproduces separately the observed trend of increasing $\dot M_\star/M_\star$ with $z$ \citep{Furlong15b}, and of decreasing size with $z$ \citep{Furlong15a}.

The strong evolution of the surface density of star formation suggests that an increasingly large fraction of galaxies drive winds at early times: whereas at $z=0$ most galaxies have $\csigmasfr \ll \sigmasfrcrit$ and hence are not expected to drive winds (these are located in the hashed region in Fig.\ref{fig_LAsfr}), at $z\gtrapprox 6$ most galaxies have $\csigmasfr \gtrapprox \sigmasfrcrit$ and their star-forming regions are expected to driven outflows. 

A consequence of the rapid increase of $\csigmasfr$ with $z$ is that $f_{\rm esc}$ also increases rapidly with $z$, as shown in Fig.~\ref{fig_fesc}. Large values of $f_{\rm esc}$ are extremely rare at $z=0$; at $z=3$ only the very brightest galaxies have significant non-zero values of $f_{\rm esc}\gtrsim 10$~per cent (in the absence of dust); but above $z=6$ galaxies with $L_{1500}>10^{28}$~erg~s$^{-1}$~Hz$^{-1}$ have a mean $f_{\rm esc}>10$~per cent, which increases further to $f_{\rm esc}>15$~per cent by $z=8$. There is good agreement (better than 10~per cent) in the predicted value of $f_{\rm esc}$ between models Ref-L0100N1504 and Recal-L0025N0752, that differ by a factor of 8 in mass resolution, for galaxies brighter than $10^{28}{\rm erg}~{\rm s}^{-1}~{\rm Hz}^{-1}$. We use the higher resolution simulation to calculate $f_{\rm esc}$ for galaxies brighter than $10^{27}{\rm erg}~{\rm s}^{-1}~{\rm Hz}^{-1}$.

Given the good agreement in the evolution of $\csigmasfr$ between \eagle\ and observed galaxies, we conclude that the increased activity in galaxies, coupled with their smaller sizes, implies that $f_{\rm esc}$ increases rapidly with redshift in observed galaxies as well. Such vigorously star forming galaxies are rare in the local Universe, but they do occur, both in the data and in the simulations which have a long tail to high $\csigmasfr$ as shown by the blue histogram in Fig.\ref{fig_LAsfr}. In fact, the blue cross  and plus symbols correspond to the low-$z$ galaxies identified by \cite{Borthakur14} and \cite{Izotov16} as having a high escape fraction corrected for dust of 20 and 8~per cent, respectively. (High redshift star-forming galaxies show no evidence of the presence of dust, \citealt{Bouwens14}.) These are located in a similar region in the $\dot M_\star-\csigmasfr$ plot of Fig.~1 as most of the currently detected brighter galaxies at $z\gtrsim 6$, and so we expect the latter to have similarly high values of $f_{\rm esc}$ . We investigate the consequences of a high escape fractions for galaxies with high values of $\sigmasfr$ next.

\vspace{-0.35cm}
\subsection{The contribution of observed galaxies to reionisation}
\begin{figure}
 \centering
 \includegraphics[width=\columnwidth]{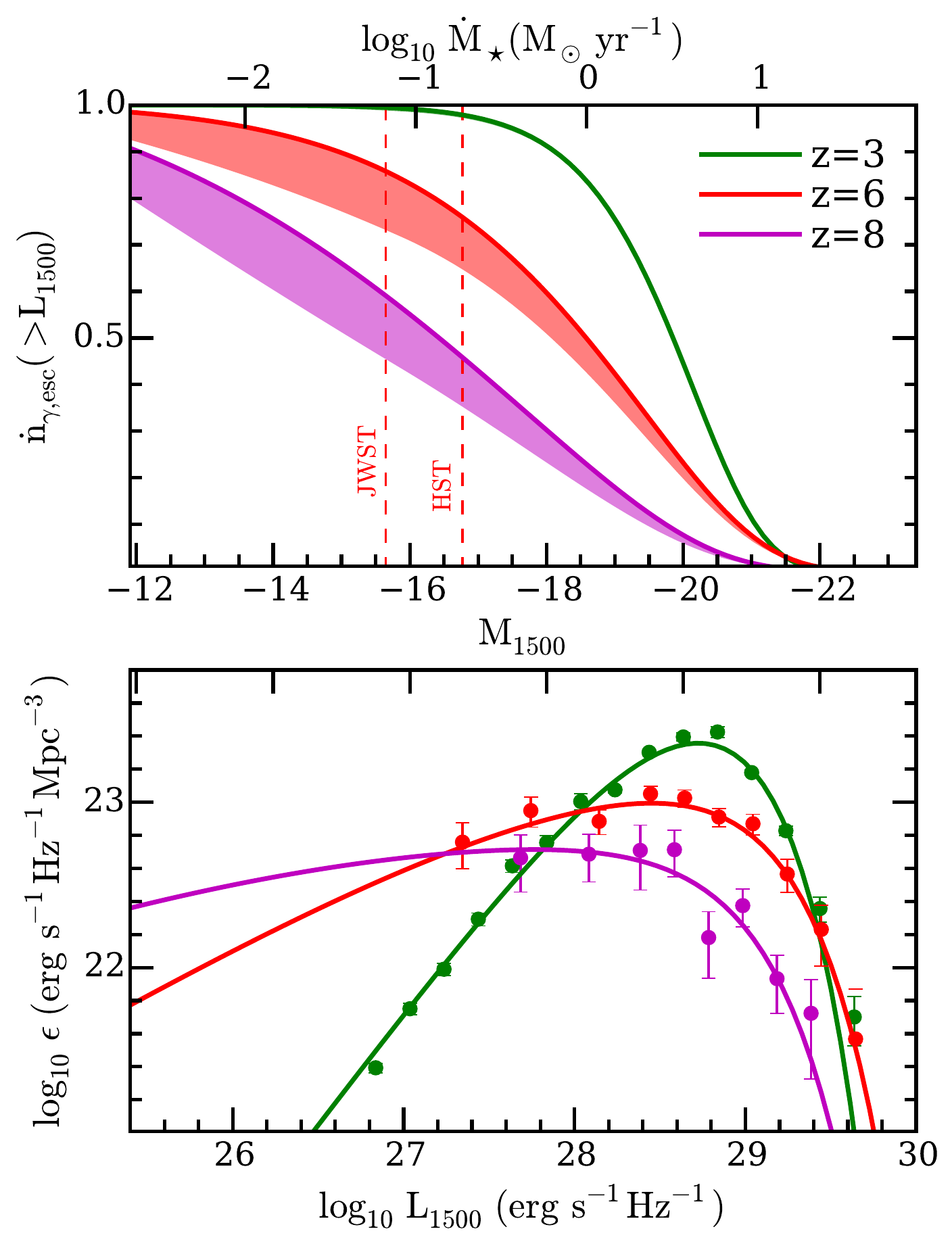}
 \caption{Detectability of the galaxies that reionised the Universe. {\bf Bottom panel:} Emissivity at 912\AA~ ($\epsilon$) as a function of the 1500~\AA\ luminosity of galaxies (middle axis shows corresponding $M_{1500}$ magnitude on the AB-system, top axis the corresponding star formation rate) for redshifts $z=3$ (green), $6$ (red) and $8$ (magenta). Curves combine the observed luminosity function of galaxies from \protect\cite{Bouwens15,Parsa15}, with model for the escape fraction of ionising photons described in the text. Even though the galaxy luminosity functions steepen rapidly towards higher redshifts, brighter galaxies dominate the emissivity even at $z=8$. {\bf Top panel:} Cumulative contribution of galaxies to the emissivity of ionising photons ($\dot n_{\rm \gamma,esc}$) at the corresponding redshifts; the HST detection limit at $z=6$ \protect\citep{Bouwens15} is shown as a red dashed line, the JWST detection limit for a 30~h integration is also indicated. { The lower edge of the shaded region corresponds to using a constant value for $f_{\rm esc}$ for all galaxies fainter than $M_{1500}=-16$}, and extrapolating the luminosity function to $M_{1500}=-10$.}
\label{fig_mnl2}
\end{figure}

\begin{figure}
 \centering
 \includegraphics[width=\columnwidth]{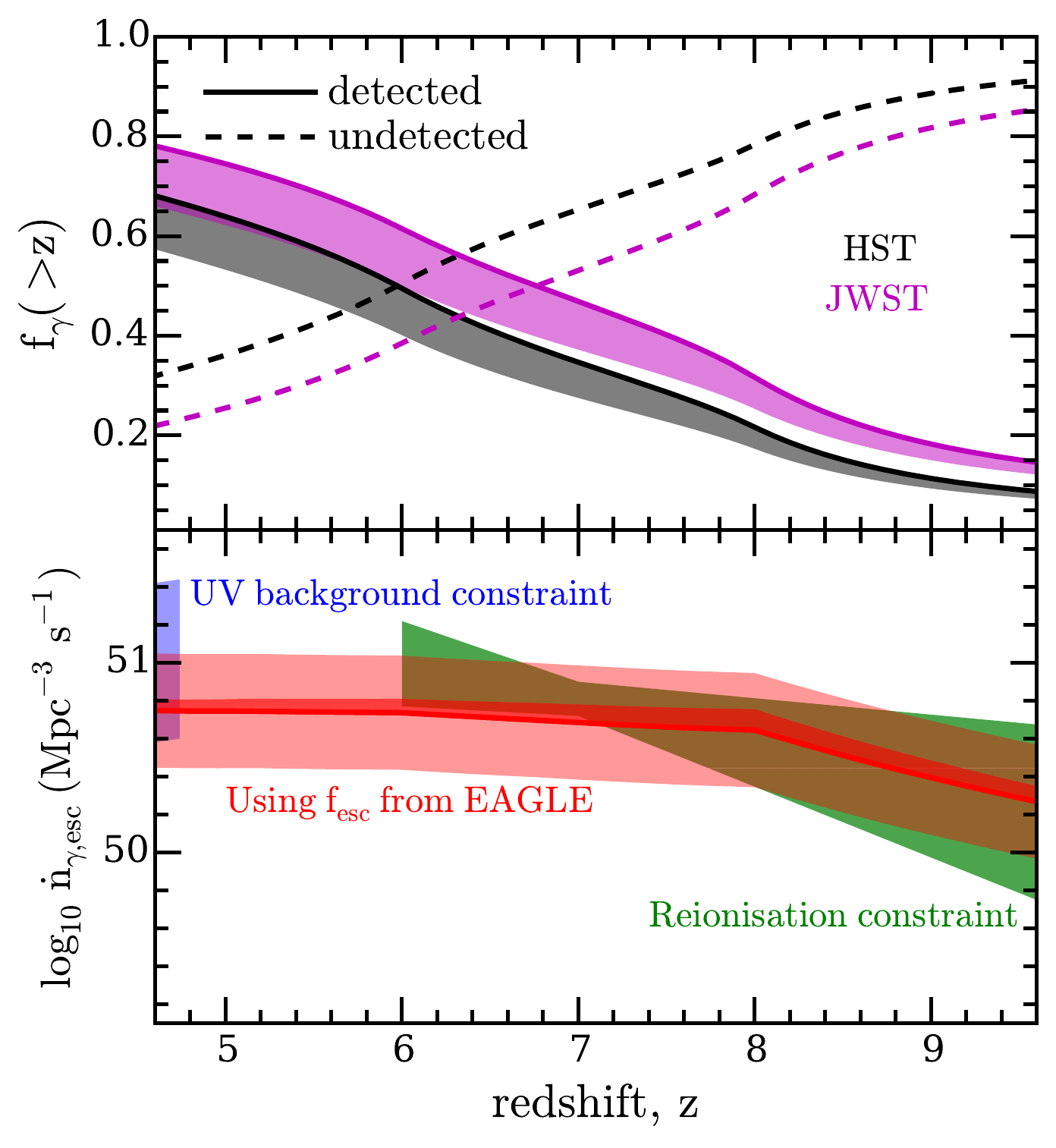}
 \caption{Total ionising emissivity ($\dot n_{\gamma, {\rm esc}}$, {\bf bottom panel}, solid red line) and fraction of that emissivity due to galaxies above a given detection limit ($f_{\gamma}(>z)$, {\bf top panel}, HST: solid black; 30~hour JWST limit: solid magenta), obtained by integrating the emissivity as function of luminosity from Fig.~\ref{fig_mnl2}. Shaded region (dark shaded region in the lower panel) shows the range if a constant value for $f_{\rm esc}$ is used for galaxies fainter than $M_{1500}=-16$, and the luminosity function is extrapolated to $M_{1500}=-10$. The light red shaded region in the bottom panel, shows the range of emissivity when the maximum allowed escape fraction is varied between 10 and 40 per cent. The {\em blue shaded region} represents the range in $\dot n_{\gamma, {\rm esc}}$ required to match the amplitude of the UV-background \citep{Becker13}; the {\em green shaded region} is the range required to match the Thomson optical depth and the evolution of the ionised volume fraction during reionisation \citep{Bouwens15}. In our model, galaxies emit enough ionising photons to match the reionisation constraints as well as the lower-$z$ amplitude of the UV-background. In this scenario, 50\% (60\%) of these photons are emitted by galaxies above the HST (JWST) limit by $z=6$ (see top panel).}
 \label{fig_mnl3}
\end{figure}

We can combine the dependence of the escape fraction on galaxy luminosity and redshift, $f_{\rm esc}(L_{1500},z)$, from Fig.~\ref{fig_fesc} with the observed number density of galaxies as function of 1500\AA\ luminosity from \cite{Bouwens15}, $\Phi(L_{1500})$, to obtain the emissivity at energies 13.6~eV, $\epsilon=f_{\rm esc}(L_{\nu},z)\,\Phi(L_{\nu},z) L(\nu_{\rm th},z)$, and the emissivity of ionising photons, $\dot n_{\gamma,{\rm esc}}(L_{1500},z)=f_{\rm esc}(L_{\nu},z)\,\Phi(L_{\nu},z)\, \int_{\nu_{\rm th}}^\infty {L_\nu\over h\nu}\,d\nu$. Here, $L_\nu$ is the luminosity of the galaxy at a frequency $\nu$ and $h\nu_{\rm th}=13.6$~eV the ionisation potential of hydrogen. We use the luminosity function at 1500 \AA\, from \cite{Bouwens15}, and assume that the intrinsic spectrum of the ionising sources has a break of  $L(912\AA)/L(1500\AA)=1/6$ as in the Starburst99 model\footnote{We use the observed luminosity function {\em without} dust correction, implicitly assuming that dust obscuration affects ionising and 1500\AA~ photons in the same way. The absolute escape fraction is then smaller than the values we quote by the (small) dust correction.} \citep{leitherer99,Bruzual03}. Resulting values of $\epsilon$ and of the cumulative contribution to the emissivity of ionising photons, $\dot n_{\rm \gamma,esc}(>L_{1500},z)$, are plotted in Fig.~\ref{fig_mnl2}.

At $z=3$, the faint-end slope of the luminosity function is sufficiently flat that bright galaxies dominate both $\epsilon$ and the photon production rate. Interestingly, even though the luminosity function becomes much steeper at $z=6$ and even more so at $z=8$ \citep{Bouwens15}, bright galaxies still dominate at these earlier times, with 50~per cent of photons escaping from galaxies brighter than $M_{1500}=-18$ at $z=6$, and $M_{1500}=-16.5$ at $z=8$. This is because the escape fraction of galaxies drops with decreasing luminosity (Fig.~\ref{fig_fesc}) faster than that the number of such galaxies increases, even when 
	luminosity function is very steep.

We show in Fig.~4 (bottom panel) that the evolution of the ionising emissivity, $\dot n_{\gamma, {\rm esc}}$, is consistent with the latest constraints on reionisation (\citealt{Bouwens15}, green band), and crucially, also with the emissivity required to produce the ionising background post-reionisation (\citealt{Becker13}, blue band). In the top we plot the cumulative fraction of ionising photons emitted up to a redshift $z$, $f_{\rm \gamma}(>z)$, by the galaxies above the current detection limit of HST (black curve) and that predicted for a 30~hour JWST integration (magenta curve). Before $z=6$, approximately $50$~per cent of the photons that reionised the Universe escape from galaxies that are above the detection limit of the {\em Hubble Ultra Deep Field}. Therefore the Universe was reionised by relatively bright, vigorously star forming, compact galaxies.

\section{Summary and conclusion}
\label{sec_CONC}
The fraction of ionising photons that escapes from galaxies ($f_{\rm esc}$ ) is a crucial ingredient in any theory of reionisation by galaxies. Observed values of $f_{\rm esc}$ at redshifts $z\lessapprox 1$ are small, $f_{\rm esc}=1-2$~per cent. This implies that, unless $f_{\rm esc}$ increases rapidly with redshift, reionisation was caused by a large population of faint galaxies, below the detection limit of the {\em Hubble Space Telescope} (HST) and possibly even below that of the {\em James Webb Space Telescope} (JWST). The existence of such a population is plausible, given the measured steep faint-end slopes of $z\gtrsim 6$ luminosity functions \cite[e.g.][]{Bouwens14}.

The escape fraction of photons from dust-free galaxies is mostly set by the structure of their interstellar medium, since ionising photons tend to be absorbed close to star forming region from which they originate. Strong winds, driven by supernovae and massive stars in a star forming region, are thought to create channels through which photons escape. Such winds tend to be observed when star formation occurs above a given surface density threshold \citep{Heckman01}, and in the few cases where photons are observed to escape in reasonable fractions of $f_{\rm esc}\gtrsim 20$~per cent, the surface density of star formation is indeed very high \citep{Borthakur14,deBarros15}.

We present a model of reionisation that encapsulates these results, by assigning a 20~per cent escape fractions to photons emerging from young star forming regions with a high value above the $\sigmasfrcrit=0.1$~M$_\odot$~yr$^{-1}$~kpc$^{-2}$ threshold of \cite{Heckman01}, and zero below this threshold. 
We computed the galaxy-averaged escape
fraction as a function of luminosity and redshift by applying this
criterion to individual gas particles in the EAGLE simulation, which
reproduces the observed star formation rate surface densities as a
function of luminosity.
We find that luminosity weighted mean value of $f_{\rm esc}$ increases rapidly with redshift, reaching values in the range 5-20~per cent at $z=6$ for galaxies above the HST detection limit, with the brighter galaxies having higher values. Combining this result with the observed luminosity function of galaxies from \cite{Bouwens15}, we obtain a model that is consistent with the latest constraints on reionisation, and on the amplitude of the ionising background post-reionisation. In this model, the {\em brighter} sources dominate reionisation.  In particular, we estimate that 50~per cent of the ionising photons that were emitted before $z=6$ originated from galaxies above the {\em Hubble Ultra Deep Field} detection limit. The instantaneous emissivity of those galaxies is $\approx 70$~per cent of the total emissivity at that redshift. JWST will be able to study these sources in detail.

\section*{Acknowledgments}
 We thank the anonymous referee for insightful comments that improved this manuscript. We gratefully acknowledge the expert high performance computing support of Lydia Heck and  Peter Draper.  We thank {\sc prace} for awarding us access to the Curie facility based in France at Tr\'es Grand Centre de Calcul. This work used the DiRAC Data Centric system at Durham University, operated by the Institute for Computational Cosmology on behalf of the STFC DiRAC HPC Facility (www.dirac.ac.uk); this equipment was funded by BIS National E-infrastructure capital grant ST/K00042X/1, STFC capital grant ST/H008519/1, STFC DiRAC Operations grant ST/K003267/1 and Durham University. DiRAC is part of the National E-Infrastructure. The study was sponsored by the Dutch National Computing Facilities Foundation, with financial support from the NWO, the ERC Grant agreements 278594 GasAroundGalaxies, GA 267291 Cosmiway, the Interuniversity Attraction Poles Programme initiated by the Belgian Science Policy Office ([AP P7/08 CHARM]), and the UK STFC (grant numbers ST/F001166/1 and ST/I000976/1). RAC is a Royal Society University Research Fellow. M.Sharma is an STFC  Post-doctoral fellow at the ICC.
\footnotesize{\bibliography{ref_eagle_He}}

\begin{thebibliography}{52}
\expandafter\ifx\csname natexlab\endcsname\relax\def\natexlab#1{#1}\fi

\bibitem[{{Becker} \& {Bolton}(2013)}]{Becker13}
{Becker} G.~D., {Bolton} J.~S., 2013, \mnras, 436, 1023

\bibitem[{{Bland-Hawthorn} \& {Maloney}(2001)}]{Bland-Hawthorn01}
{Bland-Hawthorn} J., {Maloney} P.~R., 2001, \apjl, 550, L231

\bibitem[{{Borthakur} {et~al}\mbox{.}(2014){Borthakur}, {Heckman}, {Leitherer},
  \& {Overzier}}]{Borthakur14}
{Borthakur} S., {Heckman} T.~M., {Leitherer} C., {Overzier} R.~A., 2014,
  Science, 346, 216

\bibitem[{{Bouwens} {et~al}\mbox{.}(2015{\natexlab{a}}){Bouwens},
  {Illingworth}, {Oesch}, {Caruana}, {Holwerda}, {Smit}, \&
  {Wilkins}}]{Bouwens15}
{Bouwens} R.~J., {Illingworth} G.~D., {Oesch} P.~A., {Caruana} J., {Holwerda}
  B., {Smit} R., {Wilkins} S., 2015{\natexlab{a}}, ArXiv e-prints : 1503.08228

\bibitem[{{Bouwens} {et~al}\mbox{.}(2011){Bouwens}, {Illingworth}, {Oesch},
  {Labb{\'e}}, {Trenti}, {van Dokkum}, {Franx}, {Stiavelli}, {Carollo},
  {Magee}, \& {Gonzalez}}]{Bouwens11}
{Bouwens} R.~J. {et~al.}, 2011, \apj, 737, 90

\bibitem[{{Bouwens} {et~al}\mbox{.}(2015{\natexlab{b}}){Bouwens},
  {Illingworth}, {Oesch}, {Trenti}, {Labb{\'e}}, {Bradley}, {Carollo}, {van
  Dokkum}, {Gonzalez}, {Holwerda}, {Franx}, {Spitler}, {Smit}, \&
  {Magee}}]{Bouwens14}
{Bouwens} R.~J. {et~al.}, 2015{\natexlab{b}}, \apj, 803, 34

\bibitem[{{Bridge} {et~al}\mbox{.}(2010){Bridge}, {Teplitz}, {Siana},
  {Scarlata}, {Conselice}, {Ferguson}, {Brown}, {Salvato}, {Rudie}, {de Mello},
  {Colbert}, {Gardner}, {Giavalisco}, \& {Armus}}]{Bridge10}
{Bridge} C.~R. {et~al.}, 2010, \apj, 720, 465

\bibitem[{{Brisbin} \& {Harwit}(2012)}]{Brisbin12}
{Brisbin} D., {Harwit} M., 2012, \apj, 750, 142

\bibitem[{{Bruzual} \& {Charlot}(2003)}]{Bruzual03}
{Bruzual} G., {Charlot} S., 2003, \mnras, 344, 1000

\bibitem[{{Chen} {et~al}\mbox{.}(2007){Chen}, {Prochaska}, \&
  {Gnedin}}]{Chen07}
{Chen} H.-W., {Prochaska} J.~X., {Gnedin} N.~Y., 2007, \apjl, 667, L125

\bibitem[{{Clarke} \& {Oey}(2002)}]{Clarke02}
{Clarke} C., {Oey} M.~S., 2002, \mnras, 337, 1299

\bibitem[{{Crain} {et~al}\mbox{.}(2015){Crain}, {Schaye}, {Bower}, {Furlong},
  {Schaller}, {Theuns}, {Dalla Vecchia}, {Frenk}, {McCarthy}, {Helly},
  {Jenkins}, {Rosas-Guevara}, {White}, \& {Trayford}}]{Crain15}
{Crain} R.~A. {et~al.}, 2015, \mnras, 450, 1937

\bibitem[{{Curtis-Lake} {et~al}\mbox{.}(2014){Curtis-Lake}, {McLure}, {Dunlop},
  {Rogers}, {Targett}, {Dekel}, {Ellis}, {Faber}, {Ferguson}, {Grogin},
  {Huang}, {Kocevski}, {Koekemoer}, {Lai}, {M{\'a}rmol-Queralt{\'o}}, \&
  {Robertson}}]{Curtis14}
{Curtis-Lake} E. {et~al.}, 2014, ArXiv e-prints : 1409.1832

\bibitem[{{de Barros} {et~al}\mbox{.}(2015){de Barros}, {Vanzella},
  {Amor{\'{\i}}n}, {Castellano}, {Siana}, {Grazian}, {Suh}, {Balestra},
  {Vignali}, {Verhamme}, {Zamorani}, {Mignoli}, {Hasinger}, {Comastri},
  {Pentericci}, {P{\'e}rez-Montero}, {Fontana}, {Giavalisco}, \&
  {Gilli}}]{deBarros15}
{de Barros} S. {et~al.}, 2015, ArXiv e-prints : 1507.06648

\bibitem[{{Ferrara} \& {Loeb}(2013)}]{Ferrara13}
{Ferrara} A., {Loeb} A., 2013, \mnras, 431, 2826

\bibitem[{{Fujita} {et~al}\mbox{.}(2003){Fujita}, {Martin}, {Mac Low}, \&
  {Abel}}]{Fujita03}
{Fujita} A., {Martin} C.~L., {Mac Low} M.-M., {Abel} T., 2003, \apj, 599, 50

\bibitem[{{Furlong} {et~al}\mbox{.}(2015{\natexlab{a}}){Furlong}, {Bower},
  {Crain}, {Schaye}, {Theuns}, {Trayford}, {Qu}, {Schaller}, {Berthet}, \&
  {Helly}}]{Furlong15b}
{Furlong} M. {et~al.}, 2015{\natexlab{a}}, ArXiv e-prints : 1510.05645

\bibitem[{{Furlong} {et~al}\mbox{.}(2015{\natexlab{b}}){Furlong}, {Bower},
  {Theuns}, {Schaye}, {Crain}, {Schaller}, {Dalla Vecchia}, {Frenk},
  {McCarthy}, {Helly}, {Jenkins}, \& {Rosas-Guevara}}]{Furlong15a}
{Furlong} M. {et~al.}, 2015{\natexlab{b}}, \mnras

\bibitem[{{Giavalisco} {et~al}\mbox{.}(1996){Giavalisco}, {Steidel}, \&
  {Macchetto}}]{Giavalisco96}
{Giavalisco} M., {Steidel} C.~C., {Macchetto} F.~D., 1996, \apj, 470, 189

\bibitem[{{Gnedin}(2000)}]{Gnedin00}
{Gnedin} N.~Y., 2000, \apj, 535, 530

\bibitem[{{Grazian} {et~al}\mbox{.}(2015){Grazian}, {Giallongo}, {Gerbasi},
  {Fiore}, {Fontana}, {Le Fevre}, {Pentericci}, {Vanzella}, {Zamorani},
  {Cassata}, {Garilli}, {Le Brun}, {Maccagni}, {Tasca}, {Thomas}, {Zucca},
  {Amorin}, {Bardelli}, {Cassara'}, {Castellano}, {Cimatti}, {Cucciati},
  {Durkalec}, {Giavalisco}, {Hathi}, {Ilbert}, {Lemaux}, {Paltani}, {Ribeiro},
  {Schaerer}, {Scodeggio}, {Sommariva}, {Talia}, {Tresse}, {Vergani}, {Bonchi},
  {Boutsia}, {Capak}, {Charlot}, {Contini}, {de la Torre}, {Dunlop},
  {Fotopoulou}, {Guaita}, {Koekemoer}, {Lopez-Sanjuan}, {Mellier}, {Merlin},
  {Paris}, {Pforr}, {Pilo}, {Santini}, {Scoville}, {Taniguchi}, \&
  {Wang}}]{Grazian15}
{Grazian} A. {et~al.}, 2015, ArXiv e-prints : 1509.01101

\bibitem[{{Haardt} \& {Madau}(2012)}]{Haardt12}
{Haardt} F., {Madau} P., 2012, \apj, 746, 125

\bibitem[{{Heckman}(2001)}]{Heckman01}
{Heckman} T.~M., 2001, in Astronomical Society of the Pacific Conference
  Series, Vol. 240, Gas and Galaxy Evolution, {Hibbard} J.~E., {Rupen} M., {van
  Gorkom} J.~H., eds., p. 345

\bibitem[{{Heckman} {et~al}\mbox{.}(2011){Heckman}, {Borthakur}, {Overzier},
  {Kauffmann}, {Basu-Zych}, {Leitherer}, {Sembach}, {Martin}, {Rich},
  {Schiminovich}, \& {Seibert}}]{Heckman11}
{Heckman} T.~M. {et~al.}, 2011, \apj, 730, 5

\bibitem[{{Izotov} {et~al}\mbox{.}(2016){Izotov}, {Orlitova}, {Schaerer},
  {Thuan}, {Verhamme}, {Guseva}, \& {Worseck}}]{Izotov16}
{Izotov} Y.~I., {Orlitova} I., {Schaerer} D., {Thuan} T.~X., {Verhamme} A.,
  {Guseva} N., {Worseck} G., 2016, Nature, ArXiv e-prints : 1601.03068

\bibitem[{{Kennicutt}(1998)}]{Kennicutt98}
{Kennicutt}, Jr. R.~C., 1998, \araa, 36, 189

\bibitem[{{Khaire} {et~al}\mbox{.}(2015){Khaire}, {Srianand}, {Choudhury}, \&
  {Gaikwad}}]{Khaire15}
{Khaire} V., {Srianand} R., {Choudhury} T.~R., {Gaikwad} P., 2015, ArXiv
  e-prints : 1510.04700

\bibitem[{{Kimm} \& {Cen}(2014)}]{Kimm14}
{Kimm} T., {Cen} R., 2014, \apj, 788, 121

\bibitem[{{Leitherer} {et~al}\mbox{.}(1999){Leitherer}, {Schaerer}, {Goldader},
  {Delgado}, {Robert}, {Kune}, {de Mello}, {Devost}, \&
  {Heckman}}]{leitherer99}
{Leitherer} C. {et~al.}, 1999, \apjs, 123, 3

\bibitem[{{Loeb} \& {Barkana}(2001)}]{Loeb01}
{Loeb} A., {Barkana} R., 2001, \araa, 39, 19

\bibitem[{{Ma} {et~al}\mbox{.}(2015){Ma}, {Kasen}, {Hopkins},
  {Faucher-Gigu{\`e}re}, {Quataert}, {Kere{\v s}}, \& {Murray}}]{Ma15}
{Ma} X., {Kasen} D., {Hopkins} P.~F., {Faucher-Gigu{\`e}re} C.-A., {Quataert}
  E., {Kere{\v s}} D., {Murray} N., 2015, \mnras, 453, 960

\bibitem[{{McGreer} {et~al}\mbox{.}(2015){McGreer}, {Mesinger}, \&
  {D'Odorico}}]{McGreer15}
{McGreer} I.~D., {Mesinger} A., {D'Odorico} V., 2015, \mnras, 447, 499

\bibitem[{{Mitra} {et~al}\mbox{.}(2015){Mitra}, {Choudhury}, \&
  {Ferrara}}]{Mitra15}
{Mitra} S., {Choudhury} T.~R., {Ferrara} A., 2015, ArXiv e-prints

\bibitem[{{Murray} {et~al}\mbox{.}(2011){Murray}, {M{\'e}nard}, \&
  {Thompson}}]{Murray11}
{Murray} N., {M{\'e}nard} B., {Thompson} T.~A., 2011, \apj, 735, 66

\bibitem[{{Nestor} {et~al}\mbox{.}(2013){Nestor}, {Shapley}, {Kornei},
  {Steidel}, \& {Siana}}]{Nestor13}
{Nestor} D.~B., {Shapley} A.~E., {Kornei} K.~A., {Steidel} C.~C., {Siana} B.,
  2013, \apj, 765, 47

\bibitem[{{Paardekooper} {et~al}\mbox{.}(2015){Paardekooper}, {Khochfar}, \&
  {Dalla Vecchia}}]{Paardekooper15}
{Paardekooper} J.-P., {Khochfar} S., {Dalla Vecchia} C., 2015, \mnras, 451,
  2544

\bibitem[{{Parsa} {et~al}\mbox{.}(2015){Parsa}, {Dunlop}, {McLure}, \&
  {Mortlock}}]{Parsa15}
{Parsa} S., {Dunlop} J.~S., {McLure} R.~J., {Mortlock} A., 2015, ArXiv e-prints
  : 1507.05629

\bibitem[{{Razoumov} \& {Sommer-Larsen}(2010)}]{Razoumov09}
{Razoumov} A.~O., {Sommer-Larsen} J., 2010, \apj, 710, 1239

\bibitem[{{Robertson} {et~al}\mbox{.}(2015){Robertson}, {Ellis}, {Furlanetto},
  \& {Dunlop}}]{Robertson15}
{Robertson} B.~E., {Ellis} R.~S., {Furlanetto} S.~R., {Dunlop} J.~S., 2015,
  \apjl, 802, L19

\bibitem[{{Robertson} {et~al}\mbox{.}(2013){Robertson}, {Furlanetto},
  {Schneider}, {Charlot}, {Ellis}, {Stark}, {McLure}, {Dunlop}, {Koekemoer},
  {Schenker}, {Ouchi}, {Ono}, {Curtis-Lake}, {Rogers}, {Bowler}, \&
  {Cirasuolo}}]{Robertson13}
{Robertson} B.~E. {et~al.}, 2013, \apj, 768, 71

\bibitem[{{Rutkowski} {et~al}\mbox{.}(2015){Rutkowski}, {Scarlata}, {Haardt},
  {Siana}, {Henry}, {Rafelski}, {Hayes}, {Salvato}, {Pahl}, {Mehta}, {Beck},
  {Malkan}, \& {Teplitz}}]{Rutkowski15}
{Rutkowski} M.~J. {et~al.}, 2015, ArXiv e-prints : 1511.01998

\bibitem[{{Scannapieco} {et~al}\mbox{.}(2012){Scannapieco}, {Gray}, \&
  {Pan}}]{Scannapiecco12}
{Scannapieco} E., {Gray} W.~J., {Pan} L., 2012, \apj, 746, 57

\bibitem[{{Schaye}(2004)}]{Schaye04}
{Schaye} J., 2004, \apj, 609, 667

\bibitem[{{Schaye} {et~al}\mbox{.}(2015){Schaye}, {Crain}, {Bower}, {Furlong},
  {Schaller}, {Theuns}, {Dalla Vecchia}, {Frenk}, {McCarthy}, {Helly},
  {Jenkins}, {Rosas-Guevara}, {White}, {Baes}, {Booth}, {Camps}, {Navarro},
  {Qu}, {Rahmati}, {Sawala}, {Thomas}, \& {Trayford}}]{Schaye15}
{Schaye} J. {et~al.}, 2015, \mnras, 446, 521

\bibitem[{{Schaye} \& {Dalla Vecchia}(2008)}]{Schaye08}
{Schaye} J., {Dalla Vecchia} C., 2008, \mnras, 383, 1210

\bibitem[{{Shin} {et~al}\mbox{.}(2008){Shin}, {Trac}, \& {Cen}}]{Shin08}
{Shin} M.-S., {Trac} H., {Cen} R., 2008, \apj, 681, 756

\bibitem[{{Siana} {et~al}\mbox{.}(2010){Siana}, {Teplitz}, {Ferguson}, {Brown},
  {Giavalisco}, {Dickinson}, {Chary}, {de Mello}, {Conselice}, {Bridge},
  {Gardner}, {Colbert}, \& {Scarlata}}]{Siana10}
{Siana} B. {et~al.}, 2010, \apj, 723, 241

\bibitem[{{Vanzella} {et~al}\mbox{.}(2012){Vanzella}, {Guo}, {Giavalisco},
  {Grazian}, {Castellano}, {Cristiani}, {Dickinson}, {Fontana}, {Nonino},
  {Giallongo}, {Pentericci}, {Galametz}, {Faber}, {Ferguson}, {Grogin},
  {Koekemoer}, {Newman}, \& {Siana}}]{Vanzella12}
{Vanzella} E. {et~al.}, 2012, \apj, 751, 70

\bibitem[{{Wise} \& {Abel}(2008)}]{Wise08}
{Wise} J.~H., {Abel} T., 2008, \apj, 685, 40

\bibitem[{{Wise} {et~al}\mbox{.}(2014){Wise}, {Demchenko}, {Halicek}, {Norman},
  {Turk}, {Abel}, \& {Smith}}]{Wise14}
{Wise} J.~H., {Demchenko} V.~G., {Halicek} M.~T., {Norman} M.~L., {Turk} M.~J.,
  {Abel} T., {Smith} B.~D., 2014, \mnras, 442, 2560

\bibitem[{{Yajima} {et~al}\mbox{.}(2011){Yajima}, {Choi}, \&
  {Nagamine}}]{Yajima11}
{Yajima} H., {Choi} J.-H., {Nagamine} K., 2011, \mnras, 412, 411

\bibitem[{{Zastrow} {et~al}\mbox{.}(2013){Zastrow}, {Oey}, {Veilleux}, \&
  {McDonald}}]{Zastrow13}
{Zastrow} J., {Oey} M.~S., {Veilleux} S., {McDonald} M., 2013, \apj, 779, 76

\end{thebibliography}
\end{document}